\documentclass[%
 reprint,
showpacs,
showkeys,
 amsmath,amssymb,
 aps,
]{revtex4-1}

\usepackage{graphicx}
\usepackage{dcolumn}
\usepackage{bm}

\usepackage{epstopdf}

\begin{document}

\newcommand\gsim{\buildrel > \over \sim}
\def\be{\begin{equation}}
\def\ee{\end{equation}}
{\catcode`\@=11
   \gdef\SchlangeUnter#1#2{\lower2pt\vbox{\baselineskip 0pt\lineskip0pt
   \ialign{$\m@th#1\hfil##\hfil$\crcr#2\crcr\sim\crcr}}}}
\def\gtrsim{\mathrel{\mathpalette\SchlangeUnter>}}
\def\lesssim{\mathrel{\mathpalette\SchlangeUnter<}}

\title{The Neutron Star Zoo}

\author{Alice K. Harding}
 \email{Alice.K.Harding@nasa.gov}
\affiliation{%
 Astrophysics Science Division\\
NASA Goddard Space Flight Center \\
Greenbelt, MD 20771 
}%

\date{\today}

\begin{abstract}
Neutron stars are a very diverse population, both in their observational and their physical properties.  They prefer to radiate most of their energy at X-ray and 
gamma-ray wavelengths.  But whether their emission is powered by rotation, accretion, heat, magnetic fields or nuclear reactions, they are all different species 
of the same animal whose magnetic field evolution and interior composition remain a mystery.  
This article will broadly review the properties of inhabitants of the neutron star zoo, with 
emphasis on their high-energy emission.
\end{abstract}

\maketitle
\tableofcontents

\section{Introduction}

Neutron stars are the remnants of massive stars whose cores collapse during the supernova explosions at the end of their nuclear fusion lifetimes.   Conservation of  
both the angular momentum and the magnetic flux of the progenitor star during the collapse gives the neutron star an extremely high spin rate and magnetic field.
The collapse ends when the degeneracy pressure of neutrons balances the gravitational forces of the matter.  At that point the core radius is about 10 km and 
with a mass $\sim 1.4 M_{\odot}$ the center of the star has already reached
nuclear densities.  Neutron stars thus possess the highest spin frequencies, magnetic fields and densities of any known objects in the Universe.
Since their theoretical conception by \cite{Baade1934} they have been fascinating celestial objects, both for study of their
exotic interiors and environments and for their important place in stellar evolution.  

\begin{figure*}[t]
\includegraphics[width=18cm]{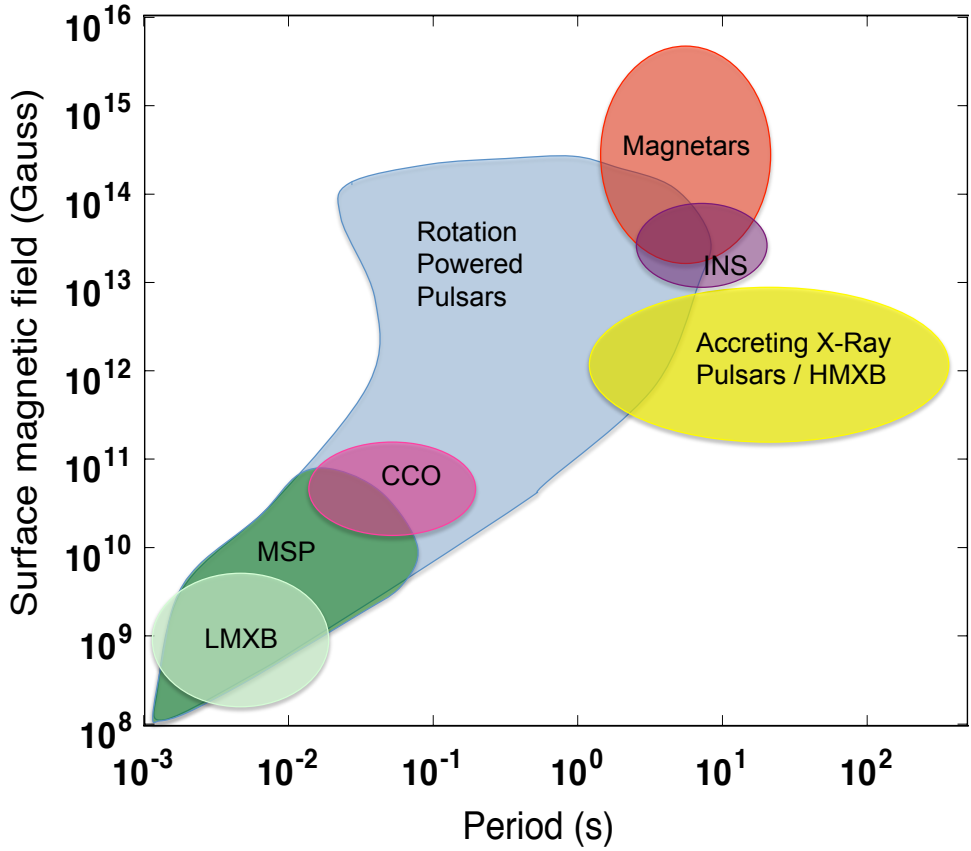}
\caption{Schematic diagram of neutron star populations with respect to their periods and derived surface magnetic field strengths.}
\label{fig:NStypes}
\end{figure*}

Neutron stars (NSs) all typically have detectable pulsations, since they are rapidly spinning and their emission patterns are highly anisotropic.  This may be one of 
the few observational properties that they have in common, because they otherwise show an amazing variety of pulsating and bursting behaviors.   
Neutron star types are classified according to the primary power source for their emission and spin evolution.  Rotation-powered pulsars (RPP) 
derive their energy primarily from the rotation 
of the NS, magnetars from magnetic field energy, isolated NSs (INS) from the latent heat of the NS matter, and accretion-powered NSs from the energy released by matter 
accreting onto the NS from a binary companion.  A subclass of accreting NSs are X-ray bursters whose bursts are powered by thermo-nuclear explosions.  An additional 
class, Central Compact Objects (CCO), are seen as soft X-ray point sources inside supernova remnants 
and seem to be quiet at all other wavelengths.  Figure \ref{fig:NStypes} shows where these different NS types roughly fall in period and surface magnetic field space.  Although period 
is a measured quantity, surface magnetic field is derived in different ways for the various NS types.  Magnetic fields in RPPs, magnetars, INS and CCO are derived from 
measured period derivatives, $\dot P$, assuming magnetic dipole radiation spin down, although in magnetars this can only give an approximate value since their spin behavior 
is complicated by magnetically-driven NS heating and bursting.  The surface dipole field at the pole determined assuming the observed $\dot P$ is 
from magnetic dipole radiation is
\be  \label{eq:Bdipole}
B_d  = \left( {\frac{{3Ic^3 P\dot P}}{{2\pi ^2 R^6 }}} \right)^{1/2} 
\simeq 2 \times 10^{12}{\rm G}\,(P \dot P_{15})^{1/2},
\ee
where $\dot P_{15} \equiv \dot P/(10^{-15}\,\rm s\,s^{-1})$, 
$P$ is in units of seconds, and $I$ ($\simeq 10^{45}$~g~cm$^2$) and $R$ ($\simeq 10^6$~cm) 
are the neutron star moment of inertia and radius.
Magnetic fields of accretion-powered NSs cannot be measured from their $\dot P$ since their spin
evolution is governed by accretion torques.  For accreting X-ray pulsars, the cyclotron lines seen in their spectra give good estimates of surface field strength:
\be \label{eq:Bcyc}
B_{\rm cyc} \simeq \left({E_c\over 11.6 {\rm keV}}\right)\,10^{12}\,{\rm G}
\ee
where $E_c$ is the cyclotron line energy.
For low-mass X-ray binaries (LMXB) and bursting sources, the Alfven radius where the NS magnetic pressure balances 
that of the accretion flow gives an estimate of the surface magnetic 
field strength:
\be \label{eq:Balfven}
B_{\rm A} \sim 10^{12}\,{\rm G}\,P_{\rm eq}\,\left({\dot M\over 10^{-9}M_{\odot}\,{\rm yr^{-1}}}\right)^{1/2}
\ee
\cite{Ghosh1979} where $P_{\rm eq}$ (in seconds) is the equilibrium spin period of the NS and $\dot M$ is the mass accretion rate.
The NS types generally occupy different areas of $P-B$ space, although there is a good deal of overlap.  The LMXB population, being progenitors of 
rotation-powered MSPs, overlap the MSP population.  CCOs lie at the lower end of the RPPs, coinciding with some of the MSP population.  The lower end of the 
magnetar population overlaps the upper end of the RPPs, and the INS overlaps both.  The accreting X-ray pulsars have magnetic fields and periods 
similar to some of the older RPPs.  The magnetars have the highest surface 
magnetic fields, while the rotation-powered millisecond pulsars and LMXBs have very low surface fields but the shortest periods.  

\begin{figure*}[t]
\includegraphics[width=18cm]{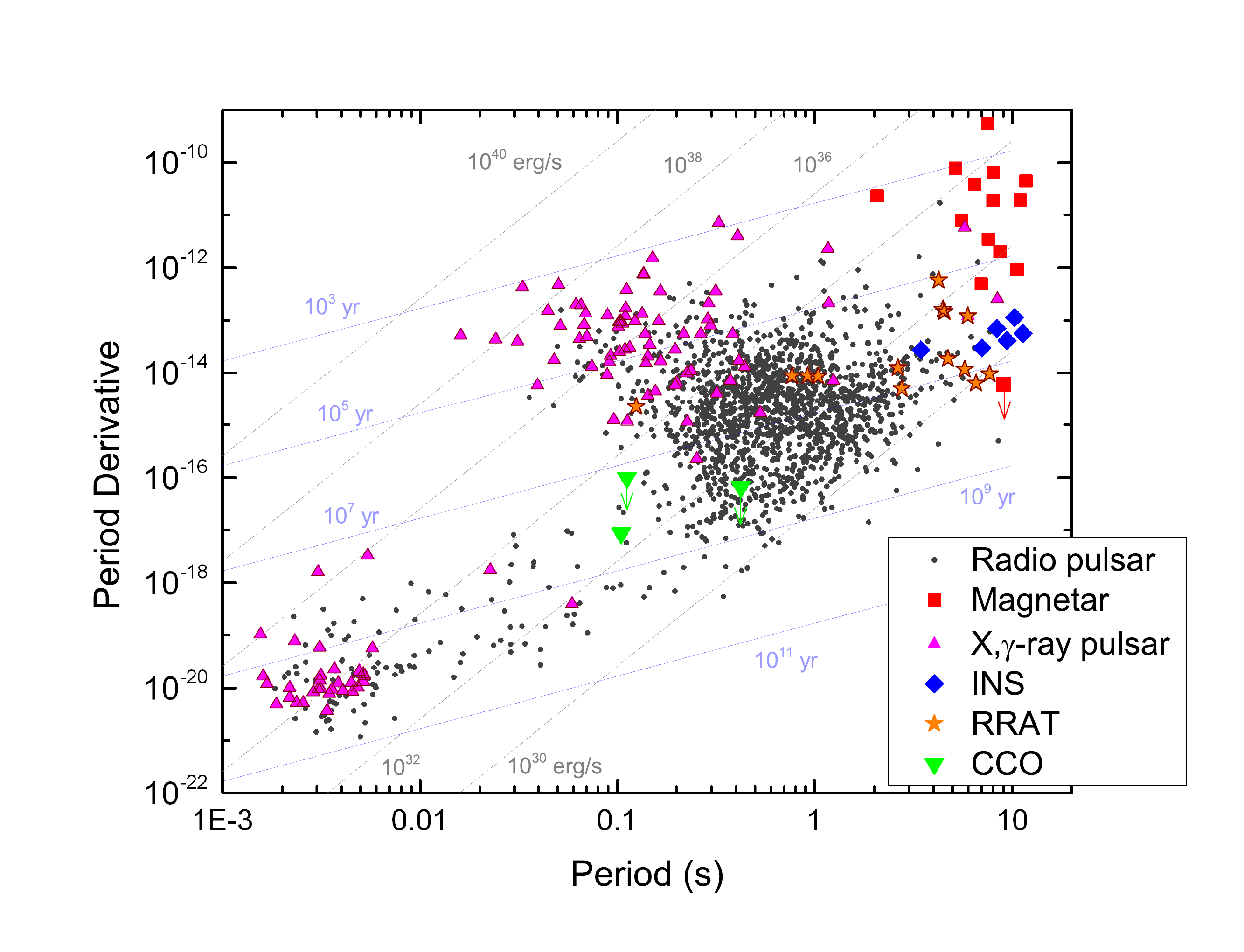}
\caption{Plot of period vs. period derivative for the presently known rotation-powered pulsars, Isolated Neutron Stars (INS), Compact Central Objects (CCO), 
Rotating Radio Transients (RRATs) and magnetars (from http://www.atnf.csiro.au/people/pulsar/psrcat/).
Lines of constant characteristic age, $P/2\dot P$, and dipole spin-down luminosity, $\dot E_d$, are superposed.}
\label{fig:PPdot}
\end{figure*}

Figure \ref{fig:PPdot} shows the distribution of non-accreting NS in measured period and period derivative.  The radio pulsars occupy the largest region of 
this phase space with their population extending from the very short period, low $\dot P$ MSPs up to the high $\dot P$, high magnetic field pulsars 
that border the magnetar range.  The magnetars have the highest $\dot P$ and some of the longest periods in the NS zoo. with some having $P \sim 11$ s
(the longest period of a radio pulsar is about 8 s).  The INS have periods very similar to those of magnetars, also reaching up to 11 s, but with $\dot P$ and 
magnetic fields about a factor of ten lower than those of the lower field magnetars.  The CCOs have very low $\dot P$, almost as low as the MSPs, 
but their spin periods are more like those of young RPPs.  High energy pulsars (RPPs with X-ray or gamma-ray pulsations) 
typically have the highest spin-down power (see Eq [\ref{eq:Edot}]) but are not necessarily the youngest, as many of the very old MSPs are efficient X-ray and gamma-ray 
pulsars.

\section{Rotation-Powered Pulsars} \label{sec:RPP}

\begin{figure*}[t]
\includegraphics[width=18cm]{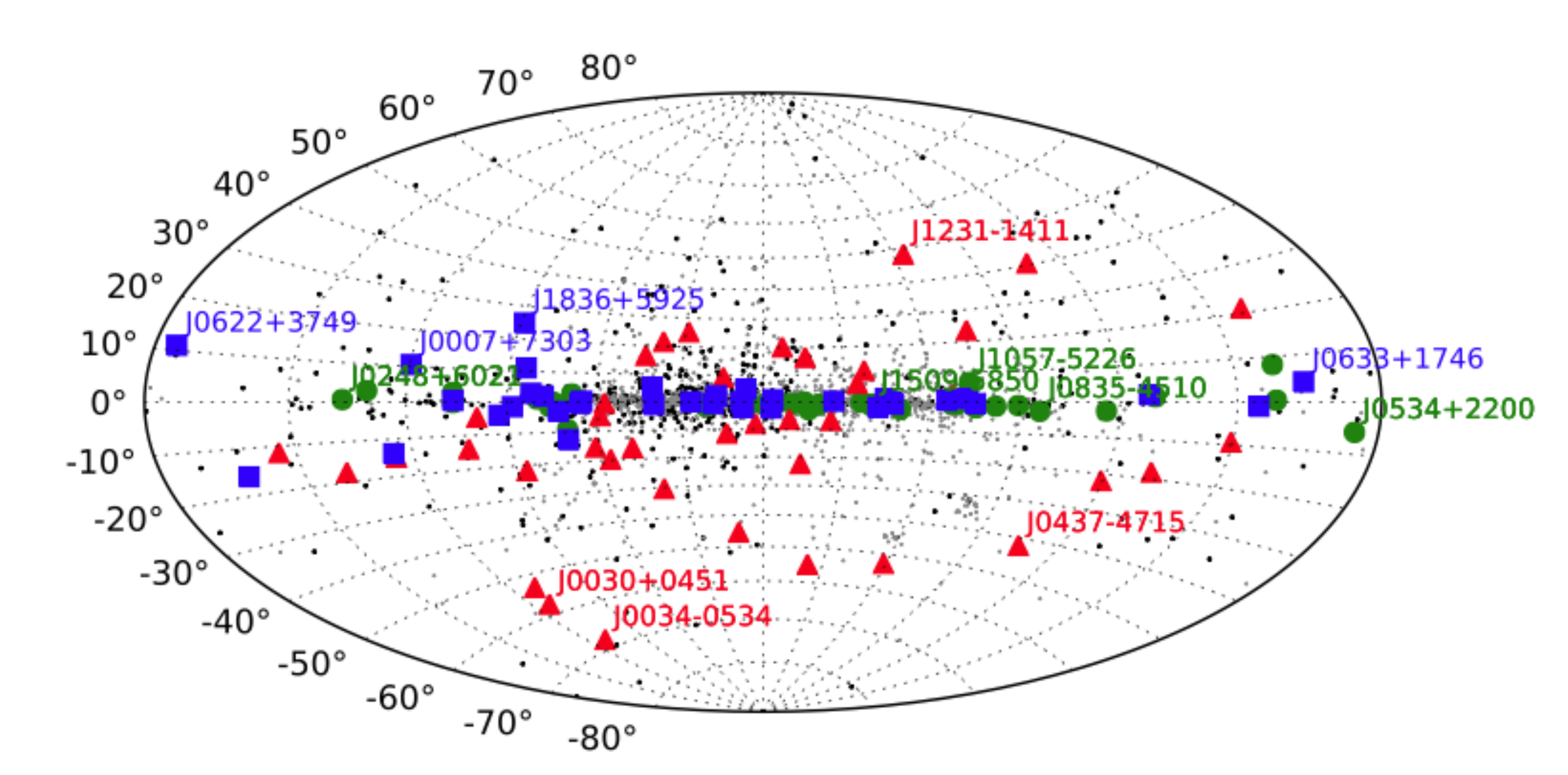}
\caption{Skymap in Galactic coordinates showing the gamma-ray pulsars of different types detected by Fermi: Blue squares: radio quiet pulsars. Red triangles: millisecond gamma-ray pulsars. Green circles: radio loud gamma-ray pulsars. Black dots: Gamma-rays were phase-folded using a rotation ephemeris. Gray dots: Pulsars
for which no rotation ephemeris was available. From \cite{Abdo2013}. }
\label{fig:2PC_Skymap}
\end{figure*}

Neutron stars that are spinning down as a result of torques from magnetic dipole radiation and particle emission are known as rotation-powered pulsars (RPP).  
The energy from their spin down appears as broad-band pulsations from radio to gamma-ray wavelengths and as a wind of energetic particles flowing into 
their surrounding pulsar wind nebulae.  Since the discovery of RPP through their radio pulsations in 1967 \cite{Hewish1968}, more than 2000 radio pulsars 
are now known with periods ranging from a few ms to several seconds \cite{Manchester2005}, http://www.atnf.csiro.au/people/pulsar/psrcat/).  
X-ray, gamma-ray and optical pulsations were soon discovered in a few of these pulsars by folding the time series obtained at these wavelengths at the radio periods.  
At present, there are over 100 RPP detected at X-ray energies and 
over 130 gamma-ray pulsars \cite{Abdo2013}; most were discovered using known radio ephemerides, but many  were also discovered through 
their X-ray or gamma-ray pulsations and are radio quiet.  The spin down of RPPs is typically smooth and predictable, but they have 
occasionally been observed to undergo sudden changes in spin called `glitches', where the period decreases and then recovers back to its normal spin-down 
rate on a timescale of days to weeks.

There are two main populations of RPPs: normal pulsars having characteristic ages $\tau = P/2\dot P$  $<$ 100 Myr, and millisecond pulsars (MSP) with 
$\tau \gtrsim 100$ Myr.  The periods of the normal pulsars are thought to have increased at a steady rate from their birth periods at the time of supernova core collapse.
The birth periods of RPPs are not well known, and inferences of the period range from tens of ms to hundreds of seconds \cite{Kaspi2005}\cite{Gotthelf2010}.  
MSPs are thought 
to have originally been members of the normal RPP population, spun down for tens of Myr and then spun up by accretion from a binary companion \cite{Alpar1982}.  
MSPs make up about 10\% of the RPP population and about 80\% of them are in binary systems.  A large number of new MSPs have recently 
been found through radio searches at positions 
of {\sl Fermi} unidentified gamma-ray sources (e.g. \cite{Ransom2011}), doubling the known numbers of radio MSPs in the Galactic disk.  
Their radio ephemerides can then be used to find the gamma-ray 
pulsations.  They are extremely good clocks since they rarely glitch, they 
have very little of the timing noise seen in young RPPs, and their spin down is very stable.  These characteristics make them potentially valuable sources for use in celestial 
navigation \cite{Coll2009} and gravitational wave detection \cite{Hobbs2010}.  

\begin{figure*}[t]
\includegraphics[width=8cm]{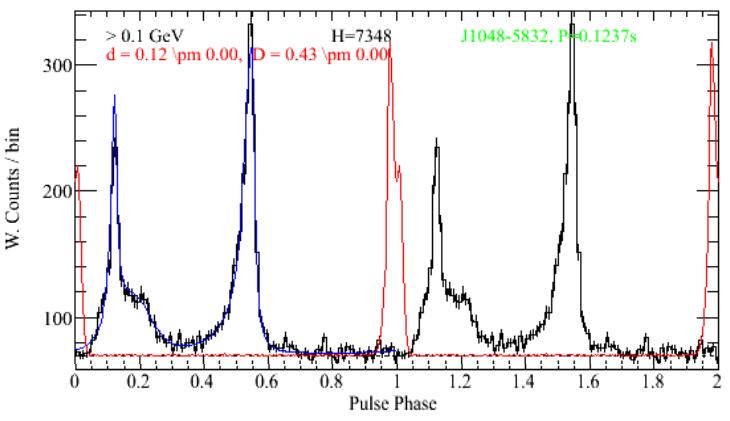}
\includegraphics[width=8cm]{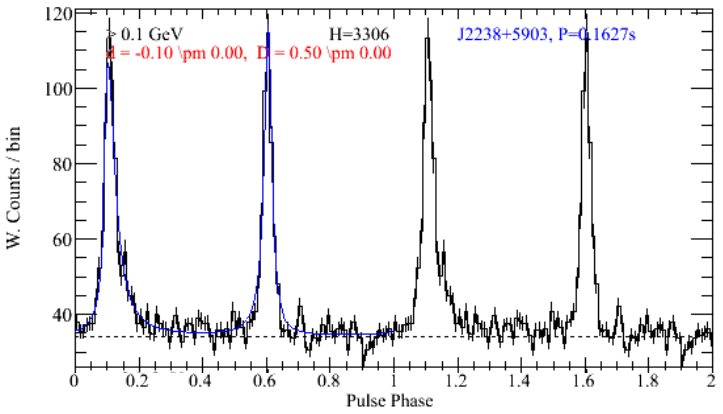}
\includegraphics[width=8cm]{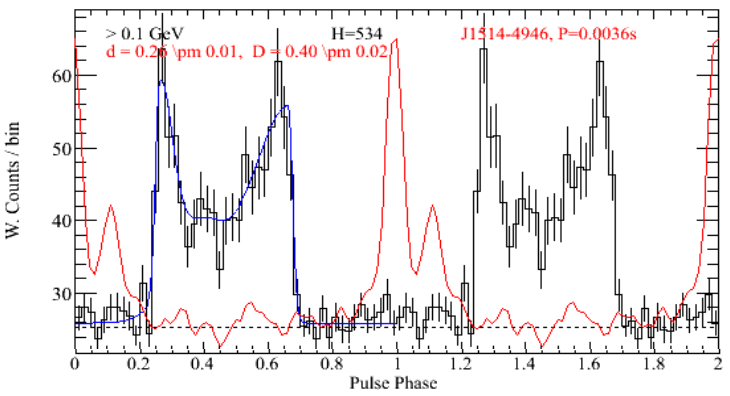}
\includegraphics[width=8cm]{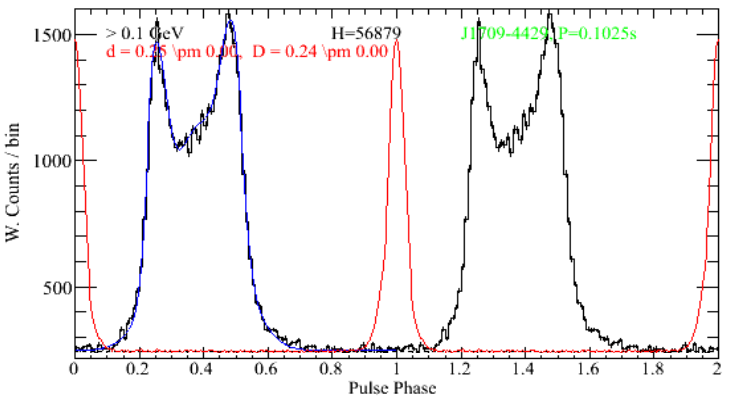}
\caption{Gamma-ray (black) and radio (red) pulse profiles for a selection of pulsars detected by {\sl Fermi}, from \cite{Abdo2013}.  }
\label{fig:FermiProfiles}
\end{figure*}

The rotating dipole model that seems to well describe RPPs gives estimates of their main electrodynamic properties.  The Poynting flux of a rotating dipole in vacuum
gives an estimate of the spin-down power:
\be \label{eq:Edot}
\dot E_d = {4\pi^2 I \dot P\over P^3} = {2B_0^2 \Omega^4 \over 3R^6 c^5} = 10^{31}\,{\rm erg\,s^{-1}}\,B_{12}^2\,P^{-4}
\ee
where $\Omega = 2\pi/P$ is the spin angular velocity, $R$ the NS radius, $B_0$ is the surface magnetic field and $B_{12} \equiv B_0/10^{12}$ G.  
The rotating dipole induces strong electric fields,
${\bf E = \Omega \times r \times B}$ that in vacuum are orders of magnitude larger than the gravitational force, pulling charges out of the NS surface to fill the magnetosphere
\cite{GJ1969}.  The maximum charge density, which screens the electric field along magnetic field lines, is 
\be
\rho_{\rm GJ} \simeq  -{{\bf \Omega \cdot B} \over 2\pi c}
\ee
which gives a scale for the current density flowing along the open field lines, $J_{\rm GJ} = \rho_{\rm GJ} c$.
The particles and magnetic field corotate with the NS out to the light cylinder radius, $R_{\rm LC} = c/\Omega$, where the corotation velocity reaches the speed of light.
The magnetosphere is divided into regions of open field lines that cross $R_{\rm LC}$ and regions of field lines that close within $R_{\rm LC}$.  The footpoints 
on the NS surface of the open/closed boundary define the polar caps.  The induced potential drop in vacuum across the open field lines is
\be
V_{\rm pc} \simeq {1\over 2} \left({2\pi \over cP}\right)^2\,B_0 R^3 = 6 \times 10^{12}\,{\rm Volts}\,B_{12}\,
P^{-2}
\ee
The open field lines and the particles that flow along them become the pulsar wind, which carries the bulk of the Poynting flux beyond  $R_{\rm LC}$.  

The pulsed radiation of RPPs carries less than about 10\% of the total spin-down power.  Most of the power in pulsed emission is in gamma rays around a GeV
for all but a few RPPs, notably the Crab pulsar whose pulsed emission power peaks in hard X rays.   The radio pulsations make up typically only $10^{-4}$ or less 
of the spin-down power, but are easier to detect since the photon flux is much higher.  Since the gamma-ray pulsations typically carry the highest percentage of 
$\dot E_d$, they can reveal the most information about the particle acceleration in pulsar magnetospheres.  The discovery of gamma-ray pulsations from over 130 
RPPs by the {\sl Fermi Gamma-Ray Space Telescope} \cite{Abdo2010, Abdo2013}, where there were only seven known previously \cite{Thompson2001}, 
has thus revolutionized the study of this type of NS (see Figure \ref{fig:2PC_Skymap}).  Two new populations of gamma-ray pulsars have been established with {\sl Fermi}: radio quiet
pulsars discovered through blind searches for gamma-ray pulsations \cite{Abdo2009b,Pletsch2012}, and gamma-ray MSPs whose pulsations are found 
using radio ephemerides \cite{Abdo2009c}.  These are presently equal in number to the radio-loud gamma-ray pulsars.
The gamma-ray pulse profiles show recurring patterns of two narrow peaks separated by phase intervals of 0.1 - 0.5, with occasional single peaks.  The most striking 
pattern is that the gamma-ray peaks are not in phase with the radio pulses, but typically arrive later in phase, with a lag that is correlated with the gamma-ray 
peak separation (smaller lags for larger peak separation, see Figure \ref{fig:FermiProfiles}).  The pulsed gamma-ray spectra are power laws with high-energy exponential 
cutoffs in the range 1 - 10 GeV.
The pulsed gamma-ray emission  of RPPs results from radiation of particles accelerated to energies of order 10 TeV by electric fields parallel to the magnetic fields
in the open magnetosphere.  Acceleration could take place near the polar caps \cite{Arons1979}, in the outer magnetosphere near the light cylinder \cite{Cheng1986} 
in slot gaps along the last open field lines
\cite{Arons1983, MH2004} or from reconnection in the striped wind outside the light cylinder \cite{Petri2005}.   The {\sl Fermi} 
measurement of an exponential shape for the spectral cutoff of the Vela pulsar \cite{Abdo2009a}
ruled out emission in the intense magnetic field near the polar caps, whose spectra would have much sharper,
super-exponential cutoffs \cite{DH1996}, as being the primary source of the observed gamma-rays.  Instead, the softer cutoffs as well as the observed profile shapes and 
peak separation vs. radio phase-lag correlation \cite{Abdo2010} predicted by outer magnetosphere models \cite{Romani1995}, strongly suggest
that the gamma-ray emission comes from high altitudes up to and possibly beyond the light cylinder.  In these models, the narrow gamma-ray peaks are caustics
formed by phase bunching of the emission along the open field boundary through aberration and light travel time effects \cite{Dyks2004,Harding2005}.  
The main radiation mechanism at GeV energies is thought to be curvature radiation by the primary particles that have been accelerated to 10 TeV energies, but inverse-Compton emission may also
contribute radiation at the higher energies (up to 400 GeV) seen from the Crab pulsar \cite{Aliu2011}.

Many RPPs (over 100 at present) also show X-ray emission, with pulsations detected in many of these.  This emission is usually made up of two components: non-thermal
emission that is probably magnetospheric, and 0.05 - 0.1 keV thermal emission from surface cooling or heated polar caps 
(see \cite{Kaspi2005} for a detailed review).  Some, like the 
Crab pulsar, show only non-thermal X-ray emission, but in these cases the thermal components are probably buried under the strong non-thermal emission.  Others, 
like MSP J2124-3358, show only thermal components with statistically insignificant non-thermal emission.  The thermal and non-thermal peaks in the pulse profiles 
are usually not in phase, but the non-thermal peaks are sometimes in phase with one or two gamma-ray peaks.  In the youngest RPPs, emission from cooling of the 
NS surface is thought to dominate the thermal radiation.  In contrast, middle-aged RPPs like Vela, Geminga and PSR J0659+1414 show a two-component thermal spectrum
from both heating and cooling, and a non-thermal power-law component.  MSPs show all of this behavior but, since these sources are too old to have 
detectable emission from cooling, the thermal components, some of which are multiple blackbodies, must be due to polar cap heating.  The non-thermal emission is in most 
models due to synchrotron radiation from electron-positron pairs, either from the polar cap or the outer gap, emitting at high altitude 
\cite{Hirotani2006, Harding2008}.  The non-cooling thermal emission could come from polar caps
heated by high-energy particles flowing toward the NS from the polar cap \cite{HM2001} or outer gap \cite{Halpern1993} accelerators.  

There is an unusual and growing subpopulation of RPPs known as Rotating Radio Transients (RRATS) (see \cite{Keane2011}, for review).  They
were discovered only very recently \cite{McLaughlin2006}, through detections of their single, isolated radio pulses.  They show a variety of transient radio behavior,
ranging from nulling (or turning off for up to $\sim 10^4$ s) for long time periods to steady pulsations that are highly modulated.  While nulling behavior in normal radio pulsars has been 
known for many years, the nulling of RRATS is extreme.  Over 70 RRATS have been identified and $\dot P$s 
have been measured for about 20 of these sources.   Their $P$ and $\dot P$ are scattered throughout the normal RPP population (see Figure 2), but they tend to be older
($\tau \gtrsim 10^5$ yr) and a number of them have higher than average surface magnetic fields.  X-ray pulsations have been detected from one RRAT, 
PSR J1819-1458 \cite{McLaughlin2007}, revealing a thermal spectrum and an absorption line at 0.5 keV, possibly due to proton cyclotron resonant scattering 
in a field of $2 \times 10^{14}$ G.  The causes of such radio transient behavior is unknown, but global changes in the currents or charge density in the pulsar 
magnetosphere has been suggested \cite{Kala2012, Li2012}.   
It is not yet clear how RRATS fit into the normal RPP population. They share some similar properties with INS (see Section \ref{sec:INS}),
but if they are a separate evolutionary group they would significantly raise the birthrate of NSs in the Galaxy \cite{Keane2008}.

In addition to pulsed emission, broadband, un-pulsed emission from radio to high-energy (TeV) gamma-rays is detected from the pulsar wind nebulae (PWNe) 
associated with RPPs.  The PWN is a repository of high-energy particles and fields from the RPP, accumulated over the pulsar's lifetime and 
trapped by the surrounding supernova shell that is moving
more slowly than the relativistic pulsar wind, or by the bow shock driven by the supersonic motion of the pulsar through the interstellar medium.  
The particles are mostly electron-positron pairs from the pair cascades \cite{DH1982}
that occur in the magnetosphere, but there 
could be some protons or positively-charged ions as well.  Most of the  Poynting flux of the RPP must at some point, 
either in the wind or at the wind termination shock, be transferred to the particles.  In the Crab nebula, 
the highest energy particles receive nearly the entire open-field voltage, as deduced from the $\sim 100$ MeV
cutoff in the synchrotron spectrum \cite{DJH1992,Arons2011}.
Many PWNe are detected at energies up to several or even tens of TeV \cite{deJager2005}.  This VHE emission is thought to be inverse-Compton emission from the 
nebular particles, scattering either their own synchrotron radiation as in the Crab PWN or, in most cases, the microwave and infrared background radiation.
Powerful gamma-ray flares from the Crab nebula that appear only in the synchrotron emission component 
have recently been observed by {\sl Fermi} \cite{Abdo2011a} and {\sl AGILE} \cite{Tavani2011}.  
Their short timescales of hours to days indicate that they are coming from a very small
region ($< 1$ arcsec) and their energies of up to several GeV challenge traditional models of acceleration in pulsar wind nebulae \cite{Buehler2012}.  

\section{Magnetars} \label{sec:Mag}

NSs whose primary power source is the tapping of energy stored in their magnetic fields are known 
as magnetars (see \cite{Woods2006}, for review).
There are two sub-classes of magnetars, Anomalous X-Ray Pulsars (AXPs) and Soft Gamma-Ray Repeaters (SGRs), 
that were thought for many years to be separate and unrelated objects.  Today, we know that SGRs and AXPs are both NSs 
possessing magnetic fields of unprecedented strength of $10^{14}-10^{16}$ G, 
and that show both steady X-ray pulsations as well as soft $\gamma$-ray bursts.  Their inferred steady
X-ray luminosities are about one hundred times higher than their spin-down luminosities, requiring a
source of power way beyond the magnetic dipole spin-down that powers RPPs.  
New high-energy components discovered in the spectra of a number
of AXPs and SGRs require non-thermal particle acceleration and look very similar to high-energy 
spectral components of young rotation-powered pulsars \cite{denHartog2008} (see Figure \ref{fig:spectra}).  

\begin{figure*}[t]
\includegraphics[width=18cm]{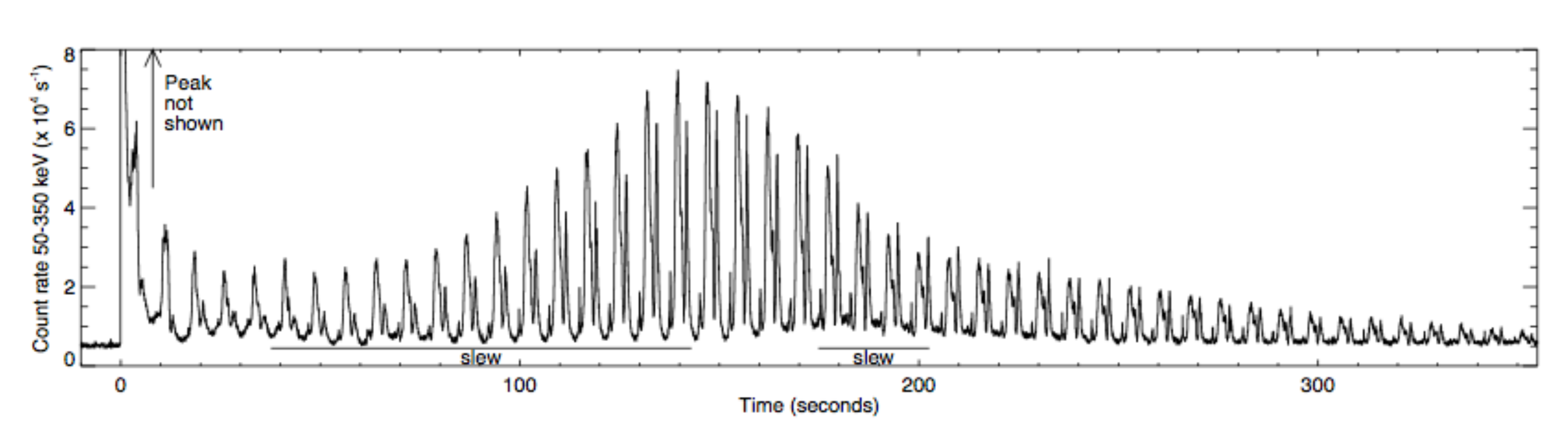}
\caption{Light curve of the 2005 superburst from SGR 1806-20, seen by the Burst Alert Telescope on Swift.  Strong 
pulsations are seen in the decaying tail of the burst. (from \cite{Palmer2005})}
\label{fig:1806}
\end{figure*}

SGRs were first detected around 1979 as $\gamma$-ray transients and were thought to be a type of 
classical $\gamma$-ray burst.  They undergo repeated bursts with several tenths of second duration and average
energy $10^{40}-10^{41}\,\rm erg$, and their bursting often occurs in episodes spaced years apart.  
They more rarely undergo giant superflares of total energy $10^{45}-10^{47}\,\rm erg$, consisting of an initial
spike of duration several hundred ms followed by a longer decay of duration several hundred seconds showing
pulsations (Figure \ref{fig:1806}).  Such superflares have been observed in three SGR sources, SGR0526-66 (the famous 5th March 
1979 event) \cite{Cline1980}, SGR1900+14 \cite{Hurley1999} and SGR1806-20 \cite{Palmer2005}.  
In 1998, SGR1806-20 was discovered to have 7.47 s pulsations in its quiescent X-ray emission \cite{Kouveliotou1998}
and a large $\dot P$ that implies a huge surface magnetic field of $\simeq 10^{15}$ G
if due to dipole spin-down.  Quiescent periodicities of 8 s and 5.16 s and large $\dot P$ were subsequently 
detected in SGR0526-66 and SGR1900+14, implying similarly high surface magnetic fields.  In all three sources, 
the quiescent periods are the same periods seen in the decay phases of their superflares.  The quiescent 
pulse profiles are very broad and undergo dramatic changes just before and after superflares.  The profiles
are often more complex, with multiple peaks before flares, changing to more simple single peaked profiles
following the flares.  Since the modulation is thought to result from beaming along magnetic field lines, the profile 
changes signal a re-arrangement of the magnetic field structure during the flares.  
All of the SGRs lie near the Galactic plane and are thought to have distances around 
10-15 kpc (except for SGR0526-66, which is in the LMC).  Recently, a transient SGR was discovered with an apparent
surface magnetic field strength $< 7.5 \times 10^{12}$ G \cite{Rea2010}.  This could be an aging magnetar
that has experienced significant field decay over its lifetime.  

The first AXPs were discovered as pulsating X-ray sources in the early 1980s by EXOSAT and were thought
to be a strange (anomalous) type of accreting X-ray pulsar.  They are
bright X-ray sources possessing luminosities (in their highest states) of $L_X \sim 10^{35}\,\rm
erg\; s^{-1}$, but show no sign of any companion or accretion disks that would be 
required to support the accretion hypothesis.  The AXPs have pulsation periods
in a relatively narrow range of 5 - 11 s and are observed to be spinning down with large period derivatives \cite{Vasisht1997}.  
Their pulse profiles are broad and very similar to those of SGR sources.
The very high surface magnetic fields of $10^{14}-10^{15}$ G implied 
by dipole spin-down were originally controversial, but have come to be accepted after the quiescent
periods were found in SGRs and especially following the recent discovery of SGR-like bursts from several AXPs \cite{Kaspi2003}.  
It is now believed that SGRs and AXPs are two varieties of the same type of object, 
very strongly magnetized, isolated NSs possibly powered by magnetic field decay.  In both sources,
the high-state quiescent luminosities of $L_X \sim 10^{35}\,\rm erg\; s^{-1}$ 
are much higher than their spin-down
luminosities of $\dot E_d \sim 2-6 \times 10^{33}\,\rm erg\; s^{-1}$, 
demanding an alternative power source.

\begin{figure*}[t]
\includegraphics[width=9cm]{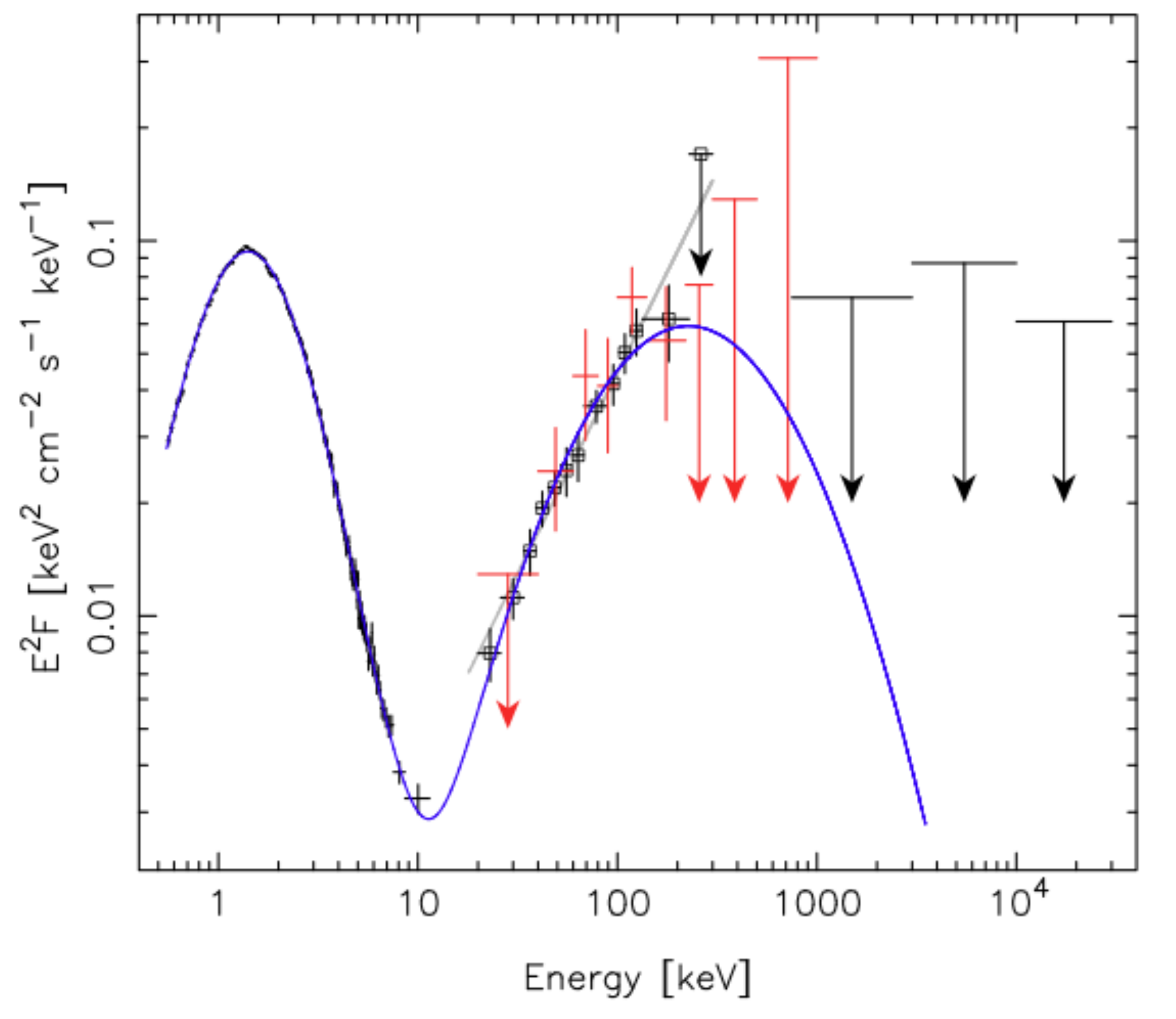}\includegraphics[width=9cm]{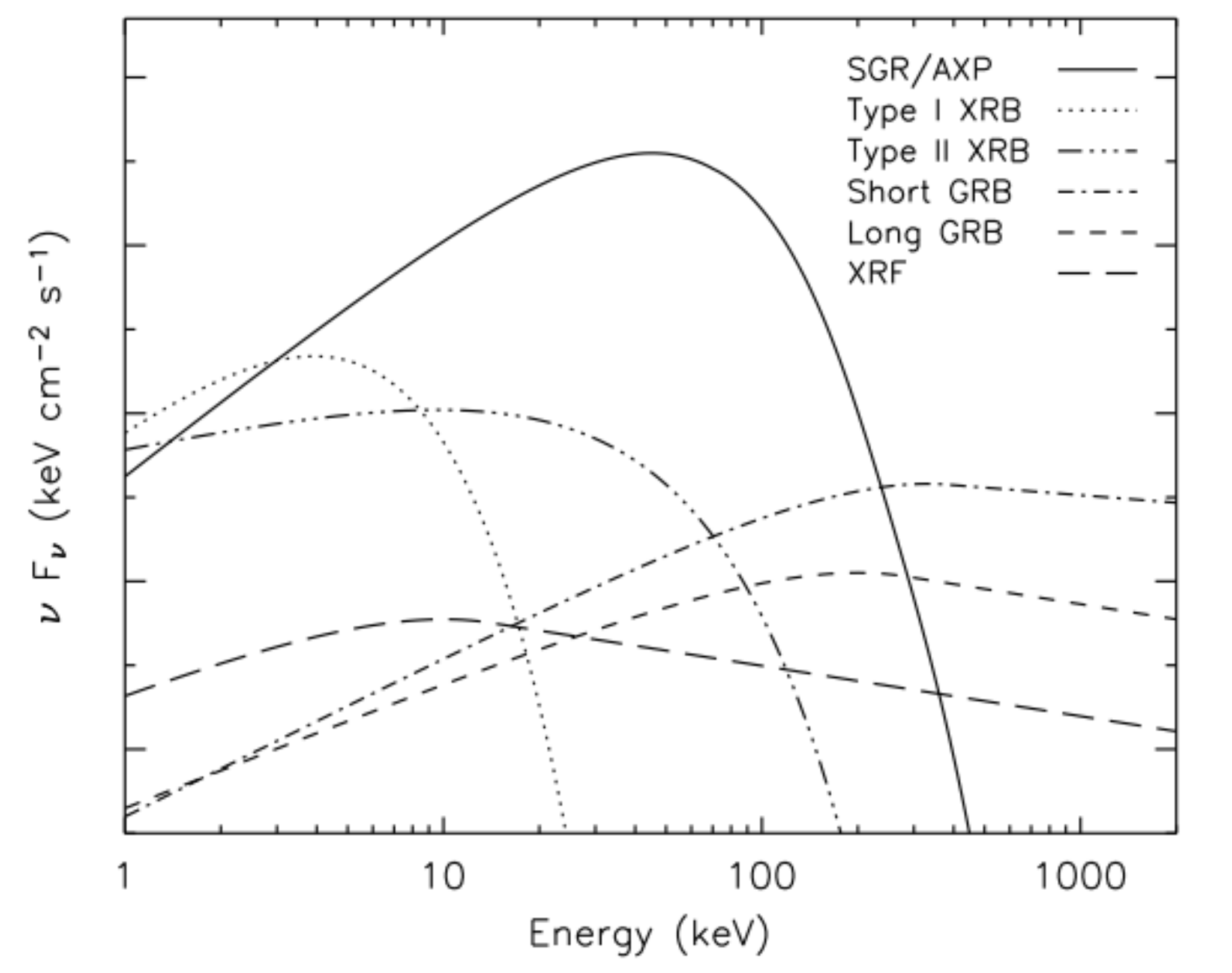}
\caption{Magnetar spectral components. Left: quiescent phase-averaged spectrum of AXP 4U 0142+61 showing lower energy 
thermal and power law components below 10 keV and high-energy component above 10 keV (from \cite{denHartog2008}).
Right: typical burst spectra from AXP/SGR, compared with burst spectra from X-ray and gamma-ray bursts (from \cite{Woods2006}).}
\label{fig:spectra}
\end{figure*}

The quiescent spectra of AXPs and SGRs (see Figure \ref{fig:spectra}) consist of a thermal component fit by $\sim$ 0.5-1 keV blackbodies
and one or more non-thermal components.  Until recently, non-thermal spectra were seen only below 10 keV and were fit with 
steep power laws having photon indices $\sim 3-4$.  When {\sl INTEGRAL} and {\sl RXTE} recently 
measured the spectra above 10 keV for the first time hard, non-thermal components were discovered in
three AXPs, and also SGR 1806-20.  In two of the AXPs, the differential
spectra between 10 keV and 50 keV are extremely flat: 1E 1841-045 \cite{Kuiper2004}
 has a power-law photon index of $s=0.94$ and
4U 0142+61 \cite{denHartog2004} displays a photon index of $s=0.45$,
both much flatter than the steep non-thermal components in the classic
X-ray band.  RXS J1708-40 possesses a slightly steeper continuum with
$s=1.18$.  The non-thermal tail of quiescent emission in SGR 1806-20 is
similarly pronounced \cite{Mereghetti2005, Molkov2005}, but
somewhat steeper, with a photon index of $s=1.6-1.9$ extending to 100 keV.  Such hard non-thermal
components require continuous particle acceleration during the quiescent state.
Transient, highly variable radio emission has been detected from several magnetars \cite{Camilo2006, 
Camilo2007} and may be correlated with their X-ray luminosity and outbursts\cite{Rea2012}.

It was proposed early on \cite{Ramaty1980, Katz1982} that SGRs were NSs with strong magnetic fields 
in the range $10^{11} - 10^{13}$ G to confine the super-Eddington burst radiation.  That SGRs and AXPs 
had much higher magnetic fields above $10^{14}$ G was proposed 
by Duncan \& Thompson [DT] \cite{Duncan1992} before their existence was observationally verified.
In this model, some NSs generate huge magnetic fields by dynamo action soon 
after their birth in supernova explosions.  Such high fields have several properties that can cause 
the NSs to behave differently from NS with lower fields.  These fields can decay on much 
shorter timescales, due to ambipolar diffusion \cite{Heyl1998}.
\be
t_{amb}^{}  \cong 10^5 yr_{} \left( {\frac{{B_{core} }}{{10^{15} G}}} \right)^{ - 2} 
\ee
Diffusion of magnetic flux out of the NS core on these timescales provides the power to magnetars in the 
DT model.  Magnetar-strength fields also apply higher stresses to the stellar crust, so that the yield
strain can exceed the crustal strength.  This property is responsible for the small SGR and AXP bursts in
the DT model \cite{Thompson1996}.  If a toroidal component of the field $B_{core} > 10^{15}$ G develops in the interior 
of the star, it can twist the external field \cite{Duncan2001}.  
Such action can cause the superflares if the twisted field lines reconnect.
Finally, due to the much faster heat transport in very strong magnetic fields, there is a greater heat flux
through the crust \cite{Heyl1998}.  Such a property may explain the much hotter surface 
temperatures of magnetars and the high quiescent X-ray emission.

Magnetar fields also produce a variety of different behavior of radiative processes (see  \cite{Duncan2000} and  
\cite{Harding&Lai2006} for reviews).  In general, in magnetic fields approaching and exceeding the quantum critical
field $B_{\rm cr} = 4.4 \times 10^{13}$ G, for radiative processes such as Compton scattering, cyclotron and
synchrotron emission and absorption, and pair production and annihilation, classical descriptions are 
largely inaccurate and QED descriptions must be used.  In addition, new processes become possible in strong
fields, such as one-photon pair production and annihilation, vacuum polarization and photon splitting, 
that cannot take place in field-free environments.  These processes, in particular vacuum polarization \cite{Lai2005}, 
strongly influence the propagation of photons in the NS atmosphere and the spectrum of the 
emergent radiation.  

According to the DT model \cite{Thompson1995}, 
the magnetar superflares result from reconnection of sheared or twisted external
field lines, leading to particle acceleration and radiation of hard emission.  The estimated luminosity of
such events,
\be
\frac{{B_{core}^2 }}{{8\pi }}R^3  \approx 4 \times 10^{46} {\rm erg} \left( {\frac{{B_{core} }}{{10^{15} G}}} \right)^2, 
\ee
is similar to observed luminosities of superflares.  The smaller bursts result from cracking of the crust,
which is continually overstressed by diffusion of magnetic flux from the NS interior.  The shaking of 
magnetic footpoints then excites Alfven waves that accelerate particles.  The energy radiated in such
events would be
\be
E_{SGR}  \cong 10^{41} {\rm erg} \left( {\frac{{B_0 }}{{10^{15} G}}} \right)^{ - 2} _{} \left( {\frac{l}{{\rm 1 km}}} \right)^2 _{} \left( {\frac{{\theta _{\max } }}{{10^{ - 3} }}} \right)^2,
\ee
where $l$ is the length scale of the displacements, $B_0$ is the crustal field and 
$\theta _{\max }$ is the yield strain of the crust.
The quiescent emission in the DT model is powered by magnetic field decay through conduction of heat from
the core.  The NS crust is heated to a temperature of 
\be
T_{crust}  \cong 1.3 \times 10^6 K_{} \left( {\frac{{T_{core} }}{{10^8 K}}} \right)^{5/9} 
\ee
where $T_{core} $ is the core temperature, and luminosity
\be
L_{\rm{x}}  \cong 6 \times 10^{35} {\rm erg s^{-1}} \left( {\frac{{B_{core} }}{{10^{16} G}}} \right)^{4.4}.
\ee

\cite{Thompson2005} have proposed that the hard, non-thermal quiescent component is due to the
creation of a strong $E_{\parallel}$ induced by twisting of field in the closed magnetosphere (see also \cite{Belo2009}), 
producing synchrotron radiation from electron acceleration at high altitude. 
An alternative model for magnetar activity and emission has been discussed by 
\cite{Heyl2005}.
The burst and quiescent radiation are a result of shocks from fast-mode plasma waves and the hard quiescent 
component is due to a pair-synchrotron cascade.  On the other hand, Baring \cite{Baring2004} and Baring \& 
Harding \cite{Baring2007} propose that resonant Compton upscattering 
of thermal X-rays by accelerated particles produces the quiescent hard emission.    

\section{Compact Central Objects} \label{sec:CCO}

\begin{figure*}[t]
\includegraphics[width=14cm]{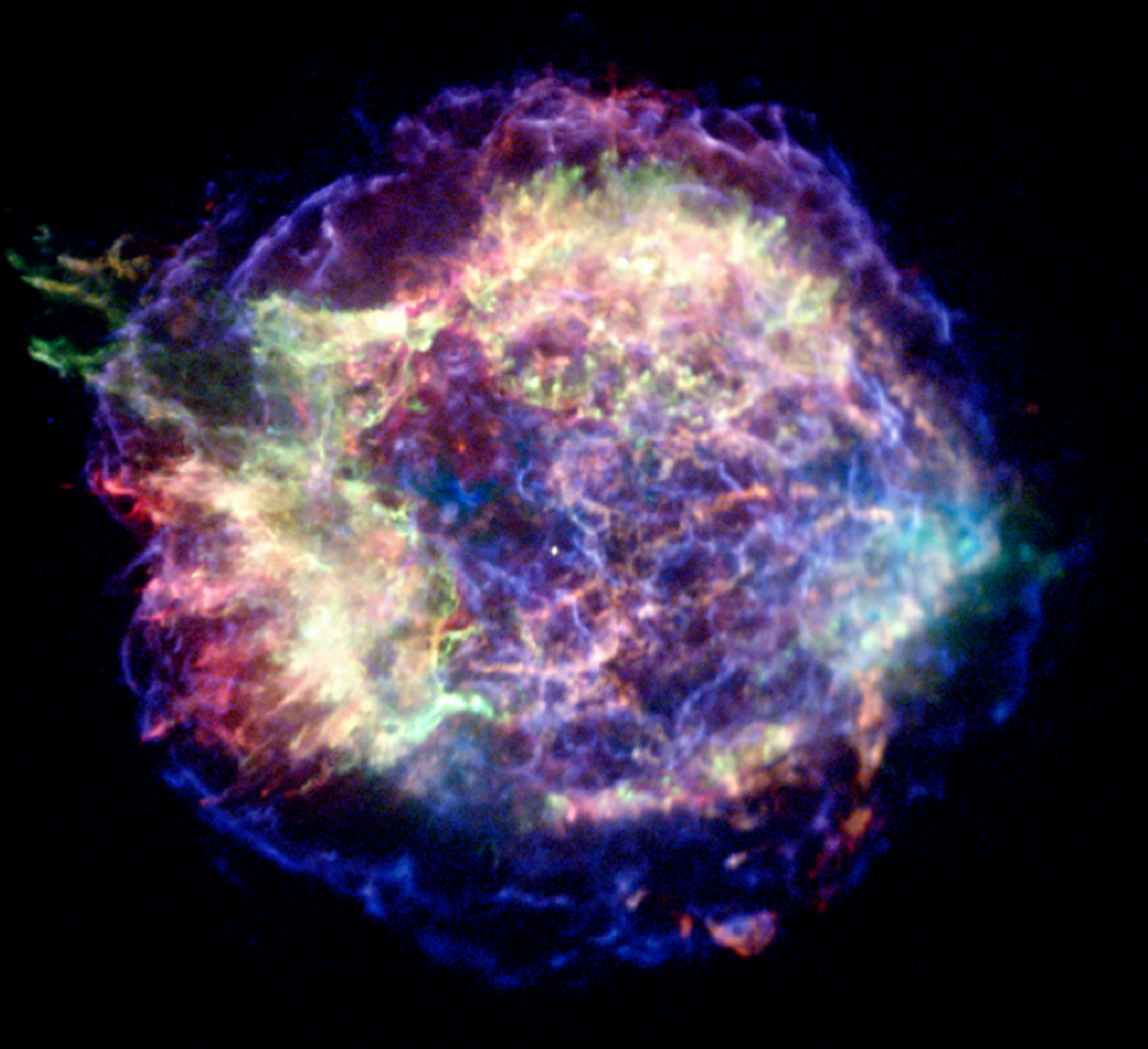}
\caption{X-ray image of the supernova remnant Cassiopeia A  taken with the Chandra X-Ray Observatory, color coded for energy 
(Red: 0.5-1.5 keV; Green: 1.5-2.5 keV; Blue 4.0-6.0 keV).  The CCO is the central 
white point source.  Credit: NASA/CXC/MIT/UMass Amherst/M.D.Stage et al. }
\label{fig:CasA}
\end{figure*}

Compact Central Objects (CCOs) are X-ray sources detected close to the centers of young supernova remnants (SNRs) that have 
no apparent emission in other wavebands and no binary companions.  Although these sources have been known and studied for several decades
without much understanding of their nature, exciting results over the past few years have brought them into the forefront of NS studies.
They have soft, exclusively thermal spectra in the few hundred eV range and X-ray luminosities around $10^{33}-10^{34}\,{\rm erg\,s^{-1}}$.  About ten CCOs are 
presently known, including the central sources of CasA (Figure \ref{fig:CasA}), Puppis A and Kes 79 supernova remnants.  Several, J1852+0040 in Kes79 
\cite{Gotthelf2010},  J0822.0-4300 in Puppis A \cite{Gotthelf2009} and 1E 1207.4 -5209 in PKS 1209-51/52 \cite{Zavlin2000} have detected pulsations in 
the hundreds of ms range.  J1852+0040 has a detected $\dot P$ \cite{Gotthelf2010}, indicating that it is spinning down like a RPP.  The measured 
$P$ and either measurements or constraints on $\dot P$  indicate that these sources have very low magnetic fields in the range $10^{10} - 10^{11}$ G
assuming magnetic dipole braking, so that their birth periods were close to their present values.  Since their SNRs are all young, $\sim 10^3 - 10^4$ yr, they were 
probably born with unusually low magnetic fields, which makes them `anti-magnetars' \cite{Gotthelf2010}.

The observed spectra are purely thermal, but some have multiple blackbody components.  The measured radii of the emitting areas, assuming 
blackbody spectra or even H atmospheres, are less that 1 km, much smaller than a NS radius.  
This is surprising if the emission is from cooling of the NS, especially since 
the very low magnetic fields do not preferentially conduct the cooling to small hotspots at the poles \cite{Zavlin1995}.  
Recently though, fits of the spectrum of the CCO in CasA with C atmosphere models, that are harder than those of H or He, give emitting radii 
that are consistent with a NS 
radius of 10 - 12 km \cite{Ho2009}.  This could be an indication of recent accretion activity that would deposit some heavy elements on
the NS surface.  Even more recently, the temperature of CasA is observed to be decreasing, by 4\% in just 10 years \cite{Heinke2010}.  
If this is in fact the first measured cooling of a NS, the rate of cooling is much more rapid than standard NS cooling models predict.  The measurements
require proton or neutron superfluid models \cite{Shternin2011} that predict sudden cooling at around the age of CasA (300 yr).
These measurements could thus also be the strongest evidence for superfluidity in NS cores.
Some CCOs show absorption lines in their spectra.  1E 1207.4 -5209 has 2-4 harmonically spaced 
lines at 0.7, 0.14, 0.21 and 0.28 keV   \cite{Bignami2003} and the spectrum of J0822.0-4300 may have an absorption 
feature at 0.8 keV \cite{Gotthelf2009}.  These could be electron cyclotron lines in a field around $8 \times 10^{10}$ G.  

Several mysteries surround these objects.  Were they really born with such low magnetic fields or were higher interior fields submerged
by mass fallback accretion to emerge later, enabling them to turn on as normal RPPs \cite{Ho2011}?  This hypothesis would fit with the need for C atmospheres 
that make their spectra consistent with full-surface NS cooling.  

\section{Isolated Neutron Stars} \label{sec:INS}

Another class of NSs that appear to be thermally cooling with no emission outside the soft X-ray band, except for faint optical/UV counterparts, 
are known as Isolated Neutron Stars (INS) (see review by \cite{Kaplan2008}).  Although these properties are similar to those of CCOs, they are a distinct 
class because they lack any observable associated supernova remnant or nebula.  There are presently seven confirmed INS (sometimes referred to as 
The Magnificent Seven), six of which have 
measured weakly modulated X-ray pulsations with periods between 3 s and 11 s, much longer than those of CCOs.  Two have high measured $\dot P$ near $10^{-13}$,
giving surface magnetic fields around $10^{13}$ G, characteristic ages $\tau \sim 2$ Myr and spin-down luminosity around $4 \times 10^{30}\,{\rm erg\,s^{-1}}$
assuming dipole spindown.  They have significant proper motions, indicating distances of less than 500 pc.
The INS thus appear to be very nearby, cooling middle-aged NSs.

The X-ray spectra of INS are soft blackbodies with kT in the range 40 - 100 eV, and X-ray luminosities in the range
$10^{30}-10^{32}\,{\rm erg\,s^{-1}}$, with no measurable power law  or non-thermal components.  Four have 
observed optical and/or UV counterparts that appear at first to be Rayleigh-Jeans extensions of the X-ray blackbodies.  However 
the optical spectra are about a factor of 10 above the extrapolation of the X-ray blackbody.   Alternatives to simple blackbody fits have been explored, 
such as a thin magnetic, ionized H atmosphere for the spectrum of J1856.5--3754, the original INS \cite{Ho2007}, or high B-field atmospheres \cite{Potekhin2012}.
Most INS show multiple or complex absorption lines in their spectra  at energy between 0.3 and 1 keV (e.g. \cite{Kaplan2009}).  It is not clear what 
causes these features, but some possibilities are proton cyclotron, and neutral or molecular H absorption.   The X-ray luminosities of several INS are larger than 
their spin down luminosities, by a factor of 60 in the case of RX J0720, so that rotation cannot power the observed X-ray emission.
Furthermore, the temperatures and thermal luminosities of INS are too high (by about a factor of 10) for conventional cooling at their characteristic ages.   This could 
imply that their magnetic fields have not been constant since birth but have decayed \cite{Heyl1998}, providing another power source in addition to cooling and making them 
younger than their spin-down ages.  

It is not clear how INS are connected to the rest of the RPP population, other than being NSs with thermal emission.  Do they lack detectable radio emission 
because our viewing angle does not cross the radio beam?    A fit of the X-ray profile of  J1856.5--3754 \cite{Ho2007} constrains magnetic inclination and 
viewing angles to be far apart.  Or are they dead radio pulsars or extreme RRATS?  They do lie very near or beyond the empirical radio death line.  But why do they not have detectable  
non-thermal high energy emission? Other RPPs of similar ages or older have detectable X-ray or gamma-ray pulsations.  
If their magnetic fields are decaying or have decayed from magnetar-strength fields, then they could be old magnetars \cite{Heyl1998} that wound up 
in the normal pulsar part of $P$-$\dot P$ space \cite{Pons2009}.  At the moment the population of INS is too small to draw any firm conclusions about their 
nature and origin, but searches for new INS are underway.

\section{Accreting Neutron Stars} \label{sec:ANS}

NSs in binary systems can accrete matter from the companion stars, either from the stellar winds or from an accretion disk that forms if the companion overflows its Roche Lobe.  
The gravitational energy from the infalling 
matter provides at least part of the energy for the observed radiation and the accretion torques dominate the spin evolution.  Despite these common properties, 
accreting NSs display a wide variety of behaviors, depending on the NS magnetic field strength, mass of the companion and properties of the accretion.

\subsection{Low-Mass X-Ray Binaries}
 
LMXBs are binary systems in which one member is a NS or black hole and the other star is a low-mass main sequence star, white dwarf or red giant that fills its Roche Lobe, transferring 
matter onto the compact object through an accretion disk.  Almost all of the radiation is emitted as X rays with a very small amount (about 1\%) in optical light.  The NS in 
LMXB are thought to be old, have weak magnetic fields ($\sim 10^{8}$ G) and are being spun up by torques from the accretion disk.  One explanation for their very weak fields
is that the accretion has reduced or submerged the magnetic flux \cite{Romani1993, Cumming2001, Ruderman2004}.   Depending on the accretion rate $\dot M$, 
they will reach an equilibrium where the pressure of the NS magnetic field balances the pressure of the accretion flow after about 1 Myr, at which point the NS period will have 
reached an equilibrium period $P_{\rm eq}$.  The value of $P_{\rm eq}$ depends on the NS field and $\dot M$ given by Eq (\ref{eq:Balfven}).  For magnetic fields around $10^8$ G
and $\dot M$ near the Eddington limit, $P_{\rm eq}$ will reach millisecond periods.  X-ray pulsations at millisecond periods were in fact found in a number of LMXBs 
with the Rossi X-Ray Timing Explorer (RXTE), 
the first being  SAX J1808.4-3658 \cite{Wijnands1998}.   Around 24 of the more than 100 LMXB sources have shown ms X-ray pulsations, with frequencies in the range 
100 - 700 Hz (see review by 
\cite{Chak2005}).  Discovery of these long-sought ms X-ray pulsations in LMXB finally established them as the progenitors of the rotation-powered MSPs.  The transition 
between an LMXB and a MSP was actually observed when ms radio pulsations were discovered from the source J102347.6+003841 which less than ten years earlier had 
shown optical emission indicating an accretion disk \cite{Arch2009}.  

Some LMXBs have steady X-ray emission, with occasional bursting behavior while others are detectable only during their bursts.  
A good fraction are transient, with outbursts on month to year timescales.  The outbursts typically last weeks to months.
The X-ray bursts are thought to be thermonuclear explosions on the NS surface, (see \cite{Strohmayer2004} for review) when the density and temperature 
of the accreted material reach the critical point for igniting nuclear reactions.
The bursts have typical timescales of around 10 - 100 s and total energy of $10^{39} - 10^{40}\,{\rm erg\,s^{-1}}$.  
LMXBs also show kHz quasiperiodic oscillations (QPO) in their X-ray emission (see \cite{vanderKlis2000} for review), thought to 
originate in the inner accretion disk flow.  In most models of QPO, the oscillation frequencies are the orbital frequencies of accreting matter.  Since these are 
stable orbits around a NS, they must be outside the innermost stable circular orbit (ISCO) determined by General Relativity, so their frequencies can be 
used to place limits on the NS mass \cite{Miller1998}.  Since the maximum measured QPO frequencies are about 1200 Hz, using the ISCO of $6GM/c^2$  
in the Schwarzschild metric gives the NS mass $M < 2 M_{\odot}$ \cite{Zhang1997}.

\begin{figure*}[t]
\includegraphics[width=16cm]{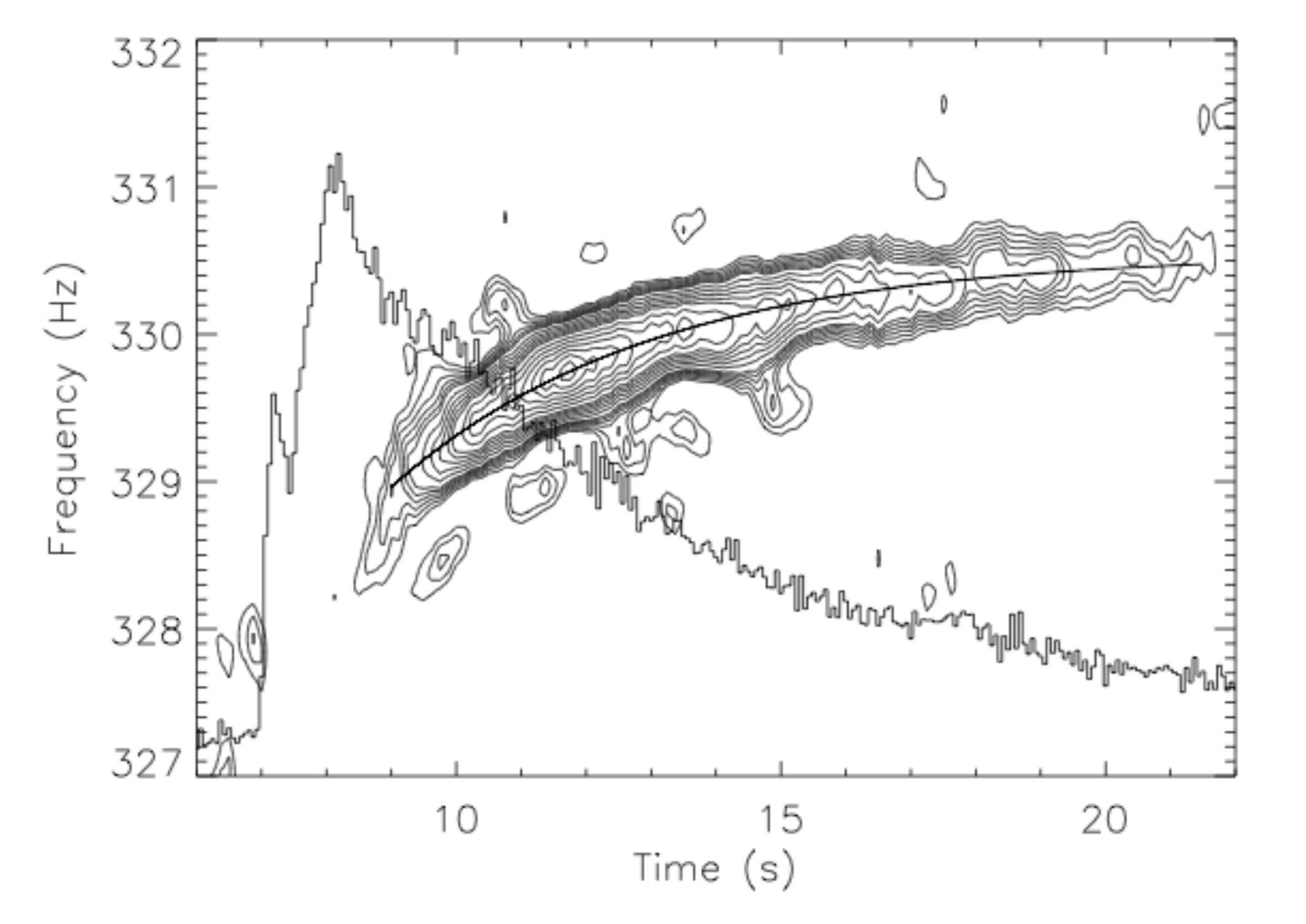}
\caption{X-ray burst oscillations in 4U 1702-43 (from \cite{Strohmayer1999}).  The histogram shows the X-ray intensity
of the X-ray burst. The contours show the Fourier power level as
a function of frequency and time, indicating a drifting oscillation starting
at 328 Hz (t = 7 s) and ending at 330.5 Hz.}
\label{fig:SM1999Fig2}
\end{figure*}

A subset ($\sim 13$) of LMXB  show ms pulsations in their persistent emission.  The rest (11) show the pulsations only during X-ray bursts \cite{Strohmayer1996, 
Watts2012} or in QPO behavior.  During the X-ray bursts, oscillations have been observed, usually during the decay phase (see Fig \ref{fig:SM1999Fig2}), 
whose frequency rises to an asymptotic limit as the intensity drops
\cite{Strohmayer1999}.  The interpretation of this behavior is that the rotation of the NS is modulating the burst hot 
spot which expands and rises in altitude, then 
drops back onto the surface.  The true rotation frequency of the NS is then the observed asymptotic frequency.  
The NS spin frequency can also appear as the frequency difference
between pairs of QPO that move up and down in frequency together.  Since the difference in the frequencies of the 
QPO pair stays the same and is in the range 200 - 400 Hz, 
it is thought to be the NS rotation frequency.  In some cases though, the frequency difference appears to be closer to 1/2 of the spin frequency, and 
the relation between QPO and spin frequency has been questioned \cite{Mendez2007}.

The distribution of NS spin frequencies from these various measurements cuts off suddenly at a maximum frequency around 620 Hz \cite{Chak2005}, 
which is well below the $\sim 1000$ Hz break-up frequency of a NS.  This indicates that there are some mechanisms that halt or counteract the spin up process.
Several possibilities have been suggested, including gravitational wave damping \cite{Harding1983,Wagoner1984}, making these good sources of kHz gravitational 
radiation for possible detection with Advanced LIGO.

\subsection{Intermediate- and High-Mass X-Ray Binaries}

Binary systems consisting of a neutron star or black hole and high-mass O or B star ($M \gtrsim 5 M_{\odot}$ ) are known as High-Mass X-Ray Binaries (HMXB), 
an example being Vela X-1.  In a large subclass of HMXBs, the NS orbits a Be star in a very eccentric orbit, accreting material only occasionally in outbursts when 
the NS crosses a disk of material surrounding the Be star.  An example of a Be X-ray binary is A0535+26.
If the donor star is an intermediate-mass star ($1.0 M_{\odot} \lesssim M \lesssim 5 M_{\odot}$), the system is an intermediate-mass X-ray binary (IMXB) \cite{Pod2002}, 
a well-known example being Her X-1.  In HMXBs, the compact object accretes material from the wind of the companion star or from a Be star disk, 
while in IMXBs the NS accretes from a disk as in LMXBs.  
Both of these classes, numbering about several hundred, are bright X-ray sources, and a large number show either 
persistent or transient X-ray pulsations (in the case of Be star binaries).  In contrast to the ms pulsations of LMXBs, pulsation 
periods of HMXBs are much longer, in the range 1 - 1000 s.  The 
relation in Eq (\ref{eq:Balfven}) then requires that the NS magnetic field in these sources are near $10^{12}$ G.  In fact, many X-ray pulsars in IMXBs and HMXBs 
(about 20) show absorption features in their spectra \cite{Heindl2004} that are almost certainly due to electron cyclotron resonance scattering.  
The energies of these cyclotron resonant scattering features (CRSF) are in the range 
10 - 50 keV, also suggesting (Eq. [\ref{eq:Bcyc}]) that the surface magnetic of the NSs are around $10^{12} - 10^{13}$ G.  

Because the NS magnetic fields are high in these systems, the Alfven radius
\be
r_{\rm A} = 3 \times 10^8\,{\rm cm}\,B_{12}^{4/7}\,\left({\dot M\over 10^{17}\,{\rm g\,s^{-1}}}\right)^{-2/7}\,\left({M/M_{\odot}}\right)^{-1/7},
\ee
lies at a much larger distance from the NS than in LMXBs, and the accreting material is strongly channeled 
via the magnetic field lines onto a small polar cap.  The infalling material then heats the NS atmosphere above the polar cap to temperatures around $10^8$ K .  The proposed 
mechanisms by which the kinetic energy of the infalling material is decelerated and transferred to heat depends on the mass accretion rate, which in turn determines its 
pressure and the source luminosity $L_{\rm X}  = GM \dot M$.  If the accretion pressure exceeds the radiation pressure, which is much higher than the spherical Eddington 
limit due to the channeling of the flow, a radiative shock will form at some distance above the NS surface \cite{Basko1976}.  
If the pressure of the accretion flow does not exceed the radiation pressure, the atmosphere is heated either by Coulomb collisions \cite{Pavlov1976} 
or a collisionless shock.  In the case where a stand-off 
shock forms, a slowly sinking accretion column radiates mostly perpendicular to the magnetic axis in a fan beam \cite{Becker2005}, while in the Coulomb-heated case, the atmosphere is a 
thin slab at the NS surface and radiates primarily along the magnetic axis \cite{Harding1984}.  The continuum spectrum of the X-ray pulsar radiation is thermal bremsstrahling and 
Compton scattering emission (e.g. \cite{Becker2007}).  The CRSFs appear as absorption lines and as many as four harmonics also appear in some X-ray pulsars such as 
4U 0115+63 \cite{Heindl1999}.  They are formed through resonant scattering of continuum photons by electrons that have a thermal distribution of momenta along the strong magnetic field lines 
(see \cite{Harding&Lai2006}, for a detailed description of the physics of CRSFs).  Since the resonant scattering cross section has a strong dependence on angle, CRSFs can provide 
a good diagnostic of the geometry and physical conditions of the emitting atmosphere \cite{Araya1999, Schonherr2007}.

\subsection{Microquasars and Gamma-Ray Binaries}

A subclass of accreting X-ray binaries are called microquasars since they display properties similar to those of quasars, including rapid variability of their X-ray emission and radio jets.  
Strong, broadband emission as well as broad emission lines  are also observed from their accretion disks. 
They consist of a compact object, either neutron star or black hole, with a normal companion.   These sources undergo repeated and sometimes periodic radio, optical and X-ray flaring 
with associated formation of relativistic jets.  Cygnus X-3 is a well-known microquasar that could be a black hole or neutron star in a 4.8 hr 
orbit around a massive hot star.  One of the strongest X-ray sources in the sky, it is also detected from radio to GeV gamma-ray wavelengths \cite{Abdo2009d}.  

The term gamma-ray binary has come to include binary systems that contain a compact object and emit high energy gamma rays.  Since several microquasars have been detected in  
gamma rays, they are one of the gamma-ray binary source types.  The other gamma-ray binary source type is a system containing a RPP orbiting a massive star.
The prototype of this source class is PSR B1259-63, which is a  48 ms radio pulsar in a 3.4 yr eccentric orbit around a Be star.  It appears as a normal radio pulsar at apastron, but at periastron 
the pulsar is thought to make several crossings through the Be star disk, producing X-rays and gamma-rays at GeV \cite{Abdo2011b} and 
TeV \cite{Ahar2005} energies.  
The X-ray binaries LSI+61 303 and LS5039 have been detected by Fermi in high-energy gamma rays \cite{Abdo2009e,Abdo2009f}, have characteristics similar to PSR B1259-63 and 
are thought to contain undetected pulsars.    The Fermi source 1FGL J1018.6Ð5856 is the first Galactic binary system to have been discovered at gamma-ray energies \cite{Abdo2012}  
but appears similar to LS5039, as both X-ray and radio modulation at the orbital period were subsequently detected. 

\section{Summary}

Neutron stars are found in a wide variety of sources, displaying an amazing array of behavior.  They can be isolated or in binary systems, accreting, heating, cooling, spinning down, spinning up, 
pulsing, flaring and bursting.  The one property that seems to determine their behavior most strongly is their magnetic field strength, structure and evolution.  The hot polar caps, bursts and flares of magnetars
are likely due to the rapid decay and twisting of their superstrong magnetic fields, whose very existence requires some kind of early dynamo activity.  The intermediate-strength magnetic fields of RPPs 
determines their spin-down behavior and radiation properties.  However, the overlap of the magnetar and RPP populations is not understood at present.  Why don't high-field RPPs burst or flare?  
Why don't lower-field magnetars sometimes behave more like RPPs? 
INS may be old magnetars whose high fields have decayed, but they do not account for the existence of younger RPPs with 
magnetar-strength fields.  Not only the strength of the magnetic field but also its configuration may be important in making a NS a magnetar or a RPP.   
Magnetic field decay is a critical link between other NS populations as well.  ``Decay" of the magnetic field is necessary for normal RPPs to evolve into MSPs
through accretion and spin up in LMXBs.  Some kind of accretion-driven field reduction is the most likely mechanism, but it is controversial since it is not clear how effective it is or on what 
timescale a buried field might re-emerge.  One piece of evidence in favor of accretion-driven field reduction is the fact that NSs in LMXBs, which are older systems ($> 10^{8}$ yr), have mostly 
low fields and NSs in HMXBs, which are younger systems ($10^7 - 10^8$ yr), have higher fields.  This may be an indication that accretion-driven field reduction or 
decay has not had enough time to operate in HMXBs but has in LMXBs.  However, there does not seem to be any evidence of decaying fields in either the LMXB or HMXB populations; e.g. smaller 
magnetic fields in older systems.   On the other hand, CCOs are very young so if they acquired their low fields through mass fallback accretion, the field 
submergence would have had to operate on much faster timescales than it apparently does in LMXBs.  But as we continue to find new species in the NS zoo, one of these may someday be the ``Rosetta Stone" that will give us the clues for solving these puzzles.

I would like to thank Megan Decesar, Isabelle Grenier, Peter Meszaros, Tod Strohmayer and Bing Zhang for providing helpful comments on the manuscript.



\nocite{*}

\begin{thebibliography}{100}

\bibitem{Abdo2009a} 
Abdo, A. A. et al. 2009a, ApJ, 696, 1084. 

\bibitem{Abdo2009b} 
Abdo, A. A. et al. 2009b, Science, 325, 840. 

\bibitem{Abdo2009c} 
Abdo, A. A. et al. 2009c, Science, 325, 848.   

\bibitem{Abdo2009d} 
Abdo, A. A. et al. 2009d, Science 326, 1512. 

\bibitem{Abdo2009e} 
Abdo, A. A. et al. 2009e, ApJ,  701, L123.  

\bibitem{Abdo2009f} 
Abdo, A. A. et al. 2009f, ApJ, 706, L56.  

\bibitem{Abdo2010} 
Abdo, A. A. et al. 2010, ApJS, 187, 460.   

\bibitem{Abdo2011a} 
Abdo, A. A. et al. 2011a, Science, 331, 739.  

\bibitem{Abdo2011b} 
Abdo, A. A. et al. 2011b, ApJ,  736, L11. 

\bibitem{Abdo2012}
Abdo, A. A. et al. 2012, Science, 335, 189. 

\bibitem{Abdo2013} 
Abdo, A. A. et al. 2013, in preparation.  

\bibitem{Ahar2005} 
Aharonian, F. A. et al. 2005, A \& A, 442, 1.

\bibitem{Aliu2011} 
Aliu, E. et al., 2011, Science, 334, 69.

\bibitem{Alpar1982} 
Alpar, M. A.; Cheng, A. F.; Ruderman, M. A.; Shaham, J. 1982, {\it Nature} {\bf 300} 28.

\bibitem{Araya1999} 
Araya, R. \& Harding, A. K. 1999, {\it ApJ} {\bf 517} 334.

\bibitem{Arch2009} 
Archibald, A. et al. 2009, Science, 324, 1411.

\bibitem{Arons1979} 
Arons, J., \& Scharlemann, E.~T. 1979, ApJ, 231, 854.

\bibitem{Arons1983} 
Arons, J. 1983, ApJ, 266, 215.

\bibitem{Arons2011} 
Arons, J. 2011, Space Sci Rev, High-Energy Emission from Pulsars and their Systems, Astrophysics and Space Science Proceedings, Springer-Verlag Berlin Heidelberg, p. 165.

\bibitem{Baade1934} 
Baade, W. \& Zwicky, F. 1934, {\it PhRv} {\bf 46} 76.

\bibitem{Baring2004} 
Baring, M.~G. 2004, in  {\it Young Neutron Stars and Their Environments}
    eds F. Camilo and B. M. Gaensler
    (IAU Symposium 218, ASP Conference Proceedings), p. 267.
    
\bibitem{Baring2007} 
Baring, M.~G. \& Harding A.~K. 2007, Astr. Space Sci., 308, 109.

\bibitem{Basko1976} 
Basko, M. M. \& Sunyaev, R. A. 1976, {\it MNRAS} {\bf 175} 395. 

\bibitem{Becker2005} 
Becker, P. A. \& Wolff, M. T. 2005, ApJ, 630, 465

\bibitem{Becker2007} 
Becker, P. A. \& Wolff, M. T. 2007, ApJ, 654, 435

\bibitem{Belo2009} 
Beloborodov, A. 2009, ApJ, 703, 1044.

\bibitem{Bignami2003} 
Bignami, G. F.; Caraveo, P. A.; De Luca, A.; Mereghetti, S. 2003, Nature, 423, 725

\bibitem{Buehler2012} 
Buehler, R. et al. 2012, ApJ, 749, 26.

\bibitem{Camilo2007} 
Camilo, F., Ransom, S. M., Halpern, J. P., \& Reynolds, J. 2007, ApJ, 666, L93 

\bibitem{Camilo2006} 
Camilo, F., Ransom, S. M., Halpern, J. P., et al. 2006, Nature, 442, 892

\bibitem{Coll2009} 
Coll, B. \& Tarantola, A. 2009, [arXiv:0905.4121].

\bibitem{Chak2005} 
Chakrabarty, D.  2005, in Binary Radio Pulsars, \ ASP Conference Series, Vol. 328, ed. F. A. Rasio and I. H. Stairs, p. 279.

\bibitem{Cheng1986} 
Cheng, K.~S., Ho, C., \& Ruderman, M.~A. 1986, ApJ, 300, 500.

\bibitem{Cline1980}
Cline, T. L. et al. 1980, ApJ Letters, 237, L1.

\bibitem{Cumming2001} 
Cumming, A., Zweibel, E., \& Bildsten, L. 2001, ApJ, 557, 958.

\bibitem{DH1982} 
Daugherty, J.~K., \& Harding, A.~K. 1982, {\it ApJ} {\bf 252} 337.

\bibitem{DH1996} 
Daugherty, J.~K., \& Harding A.~K. 1996, ApJ, 458, 278. 

\bibitem{deJager2005}
DeJager, O. C. 2005, AIP Conference Proc. 801, 298.

\bibitem{DJH1992} 
DeJager, O. C. \& Harding, A. K.  1992,  ApJ, 396, 161.

\bibitem{denHartog2004} 
denÊHartog, P.~R., Kuiper,ÊL., Hermsen,ÊW. \& Vink,ÊJ 2004, Astron. Tel. 293.

\bibitem{denHartog2008} 
den Hartog, P. R.; Kuiper, L.; Hermsen, W.; Kaspi, V. M.; Dib, R.; Knšdlseder, J.; Gavriil, F. P.  2008, A\&A, 489, 245.

\bibitem{Dyks2004} 
Dyks, J., Harding, A. K. \& Rudak, B. 2004, ApJ, 606, 1125.

\bibitem{Duncan1992} 
Duncan, R. C. \& Thompson, C. 1992, {\it ApJ}, {\bf 392}, 9.

\bibitem{Duncan2000} 
Duncan, R. C. 2000, Fifth Huntsville Gamma-Ray Burst Symposium, Eds: R. Marc Kippen, Robert S. Mallozzi, Gerald J. Fishman. AIP Vol. 526 (American Institute of Physics, Melville, New York) p.830 [astro-ph/0002442].

\bibitem{Duncan2001} 
Duncan, R. C \& Thompson, C. 2001, {\it ApJ}, {\bf 561}, 980.

\bibitem{Ghosh1979} 
Ghosh, P., \& Lamb, F. K. 1979, ApJ, 234, 296.

\bibitem{GJ1969} 
Goldreich, P., \& Julian, W.~H. 1969, ApJ, 157, 869.

\bibitem{Gotthelf2009} 
Gotthelf, E. V. \& Halpern, J. P.  2009, ApJ, 695, L35.

\bibitem{Gotthelf2010} 
Gotthelf, e. V.  \& Halpern. J. P.  2010,  ApJ, 709, 436.

\bibitem{Halpern1993} 
Halpern, J. P. \& Ruderman, M. A. 1993, ApJ, 415, 286.

\bibitem{Harding1983} 
Harding, A. K. 1983, Nature, 303, 23.

\bibitem{Harding2005} 
Harding 2005, Proc. 22nd Texas Symposium on Relativistic Astrophysics, Edited by Pisin Chen, Elliott Bloom, Greg Madejski, and Vahe Patrosian, p.149-158 [arXiv:astro-ph/0503300]

\bibitem{Harding&Lai2006} 
Harding, A. K. \& Lai, D. 2006, Rep. Prog. Phys., 69, 2631.

\bibitem{Harding1984} 
Harding, A. K., Meszaros, P., Kirk, J. G. \& Galloway, D. J. 1984, {\it ApJ}, {\bf 278} 369.

\bibitem{HM2001} 
Harding, A.~K. \& Muslimov, A.~G. 2001, ApJ, 556, 987.

\bibitem{Harding2008} 
Harding, A. K., J. V. Stern, J. Dyks \& M. Frackowiak, 2008, ApJ, 680, 1376.

\bibitem{Heindl1999} 
Heindl, W. A., Coburn, W., Gruber, D. E., et al., 1999, ApJ, 521, L49

\bibitem{Heindl2004} 
Heindl, W. A. et al. 2004, 
{\it X-ray Timing 2003: Rossi and Beyond.} AIP Conference Proceedings, Vol. 714, held 3-5 November, 2003 in Cambridge, MA. Edited by Philip Kaaret, Frederick K. Lamb, and Jean H. Swank. Melville, NY: American Institute of Physics, 2004., p.323-330

\bibitem{Heinke2010} 
Heinke, C. O. \& Ho, W. C.  2010,  ApJ, 719, L167.

\bibitem{Heyl1998} 
Heyl, J. S. \& Kulkarni, S.  R. 1998, ApJ 506, L61.

\bibitem{Heyl2005} 
Heyl, J. \& Hernquist, L. 2005, ApJ, 618, 463.

\bibitem{Hewish1968} 
Hewish, A., Bell, S. J., Pilkington, J. D., Scott, P. F. \& Collins, R. A. 1968, {\it Nature} {\bf 217} 709.

\bibitem{Hirotani2006} 
Hirotani, K.  2006, ApJ, 652, 1475.

\bibitem{Ho2011} 
Ho, W. C. 2011, MNRAS, 414, 2567

\bibitem{Ho2007} 
Ho, W. C. 2007, MNRAS, 380, 71.

\bibitem{Ho2009} 
Ho, W. C.  \& Heinke, C. O.  2009, Nature, 462, 71.

\bibitem{Hoetal2007} 
Ho W. C. G., Kaplan D. L., Chang P., van Adelsberg M., Potekhin A. Y., 2007, MNRAS, 375, 821 

\bibitem{Hobbs2010} 
Hobbs, G. et al. 2010, Classical and Quantum Gravity, 27:8.

\bibitem{Hurley1999} 
Hurley, K.  et al. 1999, ApJ, 510, L110.

\bibitem{Kala2012} 
Kalapotharakos, C., Kazanas, D., Harding, A., \& Contopoulos, I. 2012, ApJ, 749, 2

\bibitem{Kaplan2008} 
Kaplan, D. L.  2008, 40 YEARS OF PULSARS: Millisecond Pulsars, Magnetars and More. AIP Conference Proceedings, Volume 983, pp. 331-339

\bibitem{Kaplan2009} 
Kaplan, D. L. \& van Kerkwijk, M. H.  2009, ApJ, 705, 798.

\bibitem{Kaspi2003} 
Kaspi, V. et al. 2003, ApJ, 588L, 93.

\bibitem{Kaspi2005} 
Kaspi, V.M., Roberts, M.S.E., \& Harding, A.K. 2005, in Compact Stellar X-ray Sources, ed. W.H.G. Lewin \& M. van der Klis, 

\bibitem{Katz1982} 
Katz, J. I. 1982, {\it ApJ},  {\bf 260}, 371.

\bibitem{Keane2008} 
Keane E. F., Kramer M., 2008, MNRAS, 391, 2009.

\bibitem{Keane2011} 
Keane, E. F. \& McLaughlin, M. A. 2011, Bull. Astr. Soc. India, 39, 333.

\bibitem{Kouveliotou1998} 
Kouveliotou, C.  et al. 1998, Nature, 393, 235.

\bibitem{Kuiper2004} 
Kuiper,ÊL., Hermsen,ÊW. \& Mendez, 2004, ApJLett, 613, 1173.

\bibitem{Lai2005} 
Lai, D.  2005, ASTROPHYSICAL SOURCES OF HIGH ENERGY PARTICLES AND RADIATION. AIP Conference Proceedings, Volume 801, pp. 259-264.

\bibitem{Li2012} 
Li, J.,  Spitkovsky, A., \& Tchekhovskoy, A. 2012, ApJ, 746, 60.

\bibitem{Manchester2005} 
Manchester, R. N., Hobbs, G. B., Teoh, A. \& Hobbs, M., AJ, 129, 1993-2006 (2005)

\bibitem{Mendez2007} 
Mendez, M. \& Belloni, T. 2007, MNRAS 381, 790.

\bibitem{McLaughlin2006} 
McLaughlin M. A. et al., 2006, Nature, 439, 817.

\bibitem{McLaughlin2007} 
McLaughlin M. A. et al., 2007, ApJ, 670, 1307.

\bibitem{Mereghetti2005} 
Mereghetti, S., et al. 2005, A\& A  Lett., 433, L9.

\bibitem{Miller1998} 
Miller, M. C., Lamb, F. K. \& Psaltis, D. 1998, ApJ, 508, 791.

\bibitem{Molkov2005} 
Molkov, S., et al. 2005, A\& A  Lett., 433, L13.

\bibitem{MH2004} 
Muslimov, A.~G., \& Harding, A.~K. 2004, ApJ, 606, 1143

\bibitem{Palmer2005} 
Palmer, D.~M., et al. 2005, Nature, 434, 1107.

\bibitem{Pavlov1976} 
Pavlov, G. G. \& Yakovlev, D. G. 1976, {\it Sov. Phys. JETP} {\bf 43} 389.

\bibitem{Petri2005} 
Petri, J. \& Kirk, J. G. 2005, ApJ, 627, 37.

\bibitem{Pletsch2012}
Pletsch, H. J. et al. 2012, ApJ, 744, 105

\bibitem{Pod2002} 
Podsiadlowski, Ph.; Rappaport, S.; Pfahl, E. D. 2002, ApJ, 565, 1107

\bibitem{Pons2009} 
Pons et al. 2009, A \& A, 496, 207.

\bibitem{Potekhin2012} 
Potekhin, A. Y.; Suleimanov, V. F.; van Adelsberg, M.; Werner, K. 2012, [arXiv1208.6582].

\bibitem{Ramaty1980} 
Ramaty, R., Bonazzola, S., Cline, T. L., Kazanas, D., M\'{e}sz\'{a}ros, P. \& Lingenfelter, R. E. 1980, {\it Nature} {\bf 287} 122.

\bibitem{Ransom2011} 
Ransom, S. M. et al. 2011, ApJ, 727, L16.

\bibitem{Rea2010} 
Rea, N. et al. 2010, Science, 330, 944.

\bibitem{Rea2012} 
Rea, N.; Pons, JosŽ A.; Torres, D.o F.; Turolla, R.  2012, ApJ, 748, L12.

\bibitem{Romani1993}
Romani, R. 1993, in Isolated Pulsars, van Riper, K., Epstein, R., \& Ho, C., eds., 66

\bibitem{Romani1995} 
Romani, R. W. \& Yadigaroglu, I. A. 1995, ApJ, 438, 314.

\bibitem{Ruderman2004} 
Ruderman, M. 2004, in The Electromagnetic Spectrum of Neutron Stars (NATO-ASI Proceedings) 
eds. A. Baykal et al. (astro-ph/0410607).

\bibitem{Schonherr2007} 
Schonherr, G., et al. 2007, A\&A, 472, 353

\bibitem{Shternin2011} 
Shternin, Peter S.; Yakovlev, Dmitry G.; Heinke, Craig O.; Ho, Wynn C. G.; Patnaude, Daniel J.	2011, MNRAS, 412, L108

\bibitem{Strohmayer1996} 
Strohmayer, T. E., et al. 1996, ApJ, 469, L9

\bibitem{Strohmayer1999} 
Strohmayer, T. E., \& Markwardt, C. B. 1999, ApJ, 516, L81

\bibitem{Strohmayer2004} 
Strohmayer, T., \& Bildsten, L. 2004, in Compact Stellar X-Ray Sources, ed. W. H. G. Lewin \& M. van der Klis (Cambridge: Cambridge U. Press),  (astro- ph/0301544)

\bibitem{Tavani2011} 
Tavani, M. et al. 2011, Science, 331, 736.

\bibitem{Thompson2001} 
Thompson, D. J.  2001, HIGH ENERGY GAMMA-RAY ASTRONOMY, AIP Conference Proceedings, Volume 558, pp. 103-114

\bibitem{Thompson2005} 
Thompson \& Beloborodov 2005, ApJ, 634, 565.

\bibitem{Thompson1996} 
Thompson, C. \& Duncan, R. C. 1996, {\it ApJ} {\bf 473} 332.

\bibitem{Thompson1995} 
Thompson, C. \& Duncan, R. C. 1995, {\it MNRAS} {\bf 275} 255.

\bibitem{vanderKlis2000} 
van der Klis, M. 2000, ARA\& A, 38, 717

\bibitem{Vasisht1997} 
Vasisht, G. \& Gotthelf, E. V. 1997, ApJ, 486, L129.

\bibitem{Wagoner1984} 
Wagoner, R. V. 1984, ApJ, 278, 345

\bibitem{Watts2012} 
Watts, A. 2012, Annual Review of Astronomy and Astrophysics, vol. 50, p.609-640.

\bibitem{Wijnands1998} 
Wijnands, R., \& van der Klis, M., 1998, Nature, 394, 344.

\bibitem{Woods2006} 
Woods, P. M. \& Thompson, C. 2006,  in {\it Compact Stellar X-ray Sources},
eds. W.H.G. Lewin and M. van der Klis, Cambridge Astrophysics Series, No. 39. Cambridge, UK: Cambridge University Press, p. 547.  [astro-ph/0406133].

\bibitem{Zavlin1995} 
Zavlin, V. E.; Pavlov, G. G.; Shibanov, Y. A.; Ventura, J. 1995, A \& A, 297, 441

\bibitem{Zavlin2000}  
Zavlin, V. E.; Pavlov, G. G.; Sanwal, D.; TrŸmper, J.  2000, ApJ, 540, L25.

\bibitem{Zhang1997} 
Zhang, W.; Strohmayer, T. E.; Swank, J. H. 1997, ApJ, 482, L167.

\end{thebibliography}
\end{document}